\documentclass[aps,showpacs,prd]{revtex4}

\usepackage{amsmath, amssymb, bm, graphicx, graphics, color,mathrsfs,hyperref}

\newcommand{\be}{\begin{equation}}
\newcommand{\ee}{\end{equation}}
\newcommand{\ben}{\begin{eqnarray}}
\newcommand{\een}{\end{eqnarray}}

\newcommand{\cO}{{\cal O}}

\newcommand{\p}{\partial}
\newcommand{\na}{\nabla}

\newcommand{\tT}{\tilde T}

\newcommand{\ep}{\epsilon}

\newcommand{\ga}{\gamma}

\newcommand{\talpha}{{\tilde \alpha}}

\newcommand{\dA}{\delta A}
\newcommand{\dB}{\delta B}
\pacs{11.25.Tq, 04.50.Gh}

\begin{document}

\title{Viscosity bound for anisotropic superfluids with dark matter sector}
%%%%%%%%%%%%%%%%%%%%%%%%%%%%%%%%%%%%%%%%%%%%%%%%%%%%%%%%%%%%%%%%%%%%%%%%%%%%%%%%%%%%%%%%%%%%%%%%%%%%%%%%%%%%%

\author{Marek Rogatko}
\email{marek.rogatko@poczta.umcs.lublin.pl,
rogat@kft.umcs.lublin.pl}
\affiliation{Institute of Physics \protect \\
Maria Curie-Sklodowska University \protect \\
20-031 Lublin, pl.~Marii Curie-Sklodowskiej 1, Poland }
\author{Karol I. Wysoki\'nski}
\email{karol.wysokinski@umcs.pl}
\affiliation{Institute of Physics \protect \\
Maria Curie-Sklodowska University \protect \\
20-031 Lublin, pl.~Marii Curie-Sklodowskiej 1, Poland}

%%%%%%%%%%%%%%%%%%%%%%%%%%%%%%%%%%%%%%%%%%%%%%%%%%%%%%%%%%%%%%%%%%%%
\date{\today}
%\pacs{}

%%%%%%%%%%%%%%%%%%%%%%%%%%%%%%%%%%%%%%%%%%%%%%%%%%%%%%%%%%%%%%%%%%%%%%%%%%%%%%%%%%%%%%%%%%%%%%%%%%%
\begin{abstract}
The shear viscosity to the entropy density ratio $\eta/s$ of the anisotropic superfluid has been 
calculated by means of the gauge/gravity duality in the presence of the {\it dark matter} sector.
The {\it dark matter} has been described by the Yang-Mills field analogous to the one describing visible matter sector
and it is assumed to interact with the visible field with coupling constant $\alpha$. 
Close to the superfluid transition temperature ($T_c$) the analytical solution has been given up to the leading  
order in a symmetry breaking parameter and the ratio of the gravitational constant and Yang-Mils coupling. The 
tensor element of ratio $\eta/s$
remains unaffected by the {\it dark matter} for the viscosity tensor in the plane perpendicular to the
symmetry breaking direction (here $yz$). The temperature  dependence and 
the linear correction in $(1-\alpha)$ in the plane containing this direction (here $xy$) was also revealed. 
The correction linearly vanishes 
for  temperature tending to the critical one $T\rightarrow T_c$.
\end{abstract}
%%%%%%%%%%%%%%%%%%%%%%%%%%%%%%%%%%%%%%%%%%%%%%%%%%%%%%%%%%%%%%%%%%%%%%%%%%%%%%%%%%%%%%%%%%%%%%%%%%%%

\maketitle

%%%%%%%%%%%%%%%%%%%%%%%%%%%%%%%%%%%%%%%%%%%%%%%%%%%%%%%%%%%%%%%%%%%%%%%%%%%%%%%%%%%%%%%%%%%%%%%%%%%%
\section{Introduction}
The discovery of the universal lower bound of the shear viscosity to entropy density ratio
in strongly coupled field theories has been achieved by means of the gauge/gravity
correspondence and since then it is considered as one of the hallmarks of the holographic
approach, as well as, one of the most important insights of the AdS/CFT correspondence \cite{mal98}-\cite{gub98}
% was the derivation of the viscosity to entropy density ratio for all gauge theories with Einstein  gravity
into strongly coupled theories \cite{pol01}-\cite{iqb09}, with direct relevance to the laboratory physics. 
The universality~\cite{kov05} of the ratio $\eta=1/4\pi$ in units of $\hbar/k_B$ 
(the Plancks constant over the Boltzmann constant), also known as 
 the Kovtun-Son-Starinets or KSS bound, has sparked a new research wave seeking the
conditions leading to its violation. These include higher curvature corrections, Gauss-Bonnett modifications of the
gravity, the non-trivial couplings to the dilaton field, the anisotropy of the background as, e.g., spontaneous 
spatial symmetry breaking.

In the following, we shall discuss these symmetry breaking factors in some details in the theory 
supplemented with dark matter sector.
Interestingly, there were  arguments presented that the aforementioned ratio is 
satisfied for all known fluids, like superfluid helium \cite{rup07} and in particular 
the QCD quark gluon plasma \cite{son11}. 
Explicit calculations reveal that the universal value of the viscosity to entropy density ratio 
remains intact in theories with the presence of a non-zero chemical potential \cite{mas06}-\cite{mae06}.
The calculations taking into account the stringy corrections, presented in \cite{buc05}-\cite{buc08c}, 
proved that the bound~\cite{kov05} was still satisfied. In \cite{ban10} it was shown that
studies of the near-horizon geometry of the non-extremal black objects are enough to obtain the answer.
The author have found corrections to the KSS bound resulting from the coupling to higher
curvature terms as quantified by the coupling $g$ and temperature $T$ evaluated with near-horizon data. Also the coupling to the
dilaton field~\cite{mat11,reb12} changes the aforementioned
ratio, making it again dependent on the temperature of the black hole and
the value of the dilaton field at the event horizon.  
The problem of the validity of the KSS bound for the theories treated as perturbative correction 
to the ordinary Einstein one, was studied in \cite{bru09}-\cite{bru11}.

%%%%%%%%%%%%%%%%%%%%%%%%%%%%%%%%%%%%%%%%%%%%%%%%%%%%%%%%%%%%%%%%%%%%%% 
There have been found many instances when the KSS bound is violated. 
This happens if higher curvature corrections to the ordinary Einstein theory of gravity~\cite{bri08} are considered.
It turned out that Gauss-Bonnet coupling enters 
into the relation of viscosity to entropy density ratio. It was also revealed that field 
excitations in dual field theory are responsible for a superluminal propagation velocity \cite{bri08a},
when Gauss-Bonnet coupling constant is greater than $9/100$. 
The shear viscosity of the field theory with gravity duals of Gauss-Bonnet gravity with a non-trivial dilaton 
coupling were studied \cite{cai09}. It was found that the dilaton coupling has a
 non-trivial influence on the viscosity to entropy ratio. The calculations comprising AdS 
Gauss-Bonnet gravity with $F^4$ term corrections of Maxwell field \cite{cai08},
 as well as, a gravity dual to Einstein-Born-Infeld gravity \cite{cai08b}, were elaborated. 
In the case of Born-Infeld duals it was established that
 the ratio had  still the universal value, at least up to the first order of the Born-Infeld parameter. 
The review of the universality problem of the viscosity to entropy density ratio~\cite{cre11} gives account
of the earlier studies.

%%%%%%%%%%%%%%%%%%%%%%%%%%%%%%%%%%%%%%%%%%%%%%%%%%%%%%%%%%%%%%%%%%%%%%%%%%%%%%%%%%%%%%%
There have been observed a resurgence of the interests in the problem of the universality 
of the viscosity entropy density ratio in the theories taking anisotropy of the problem into account.
This is due to the relation of the universality to the shear mode transforming properties, 
i.e., transformation as a helicity two state under a rotational symmetry.
The universality holds as one is looking for $SO(3)$ symmetry because of the 
fact that shear viscosity depends on the gravitational perturbations in tensor modes of the 
$SO(3)$ in the dual gravity. Each of the tensor mode fulfills a massless scalar field equation 
and decouples from each other. This fact ensures the universality of viscosity to entropy density ratio.

On the other hand, the anisotropic symmetry broken phase is responsible for the violation 
of the aforementioned ratio. It is caused by the fact that $SO(3)$ symmetry is broken to $SO(2)$ one.
Namely, the gravitational wave nodes in the broken symmetry directions are not of the $SO(2)$-type 
and they interact with Yang-Mills gauge fields. 
Natural field theory system breaking the $SO(3)$ symmetry and also translation symmetry is the crystalline solid.
Other examples include superfluids with vector order parameters. The first case have been carefully analyzed
in~\cite{alb16} and shown that on the gravity side it is related to massive gravity or 
better Lorentz violating massive gravity. The explicit numerical and analytical calculations~\cite{alb16a} show
the the ratio is a strongly diminishing function of the dimensionless graviton mass parameter $m$ and depends on the
square ratio of the horizon to the AdS apace time radius ($r_h/L$).

%In anisotropic theories it was also claimed the non-universality of the quoted ratio. 
%Namely, the viscoelastic nature of the mechanical response in the holographic solids was the reason
%of the viscosity bound violation \cite{alb16}.
%%%%%%%%%%%%%%%%%%%%%%%%%%%%%%%%%%%%%%%%%%%%%%%%%%%%%%%%%%%%%%%%%%%%%%%%%%%%%%%%
It was found that in five-dimensional AdS/CFT theory near the critical point, the phase 
transition depends on the ratio of five dimensional gravitational constant to Yang-Mills
coupling constant. Namely, when this ratio was smaller than the critical one, the phase 
transition became o second order one \cite{erd11}.
On the contrary, for the reverse relation one obtains the first order phase transition.

The holographic description of transport phenomena in anisotropic superfluids was 
presented in~\cite{erd12}, where  the non-universality of viscosity to entropy density ratio
 was elaborated numerically. The obtained results were in agreement with  the nemetic crystal behaviors \cite{gen74}. The authors~\cite{erd12} have also analyzed exotic transport 
properties like the flexoelectric effect.
The analytic description near the critical point (close to the phase transition) was studied 
in \cite{bas12}, where the non-universality of the ratio shows up in the directions
of broken symmetry, while the ratio in the unbroken one i.e. universal in other directions.
The shear viscosity treatment in anisotropic holographic superfluids from the point of view 
of holographic renormalization group was given in \cite{oh12}. On the other hand, the gauge/gravity duality in Einstein Gauss-Bonnet gravity was implemented 
for studies the shear viscosity entropy density ratio non-universality \cite{bha15}.

%%%%%%%%%%%%%%%%%%%%%%%%%%%%%%%%%%%%%%%%%%%%%%%%%%%%%%%%%%%%%%%%%%%%%%%%%%%%%%%%%
Motivated by the AdS/CFT correspondence, the 
black hole in a theory of Einstein gravity coupled to a scalar field, including
a non-minimal Hordenski term, where gradient of scalar couples to the Einstein tensor, were
studied in \cite{fen15}. Among all it was revealed the violation of $\eta/s$ in the theory
which Lagrangian was at most linear in curvature tensor.

%%%%%%%%%%%%%%%%%%%%%%%%%%%%%%%%%%%%%%%%%%%%%%%%%%%%%%%%%%%%%%%%%%%%%%%%%%%%%%%%%%%%%%%%%
The AdS/CFT duality is a very effective tool for studying  strongly correlated systems. 
In some sense it is a sort of a calculus for which the main ingredient is the AdS spacetime.
The content of the gravity theory defines the conditions of the considered problem. The matter 
on the gravity side give us information
about its influence on the elaborated phenomena. Therefore it might be helpful in investigations of the {\it dark matter}. 
The knowledge how the {\it dark matter} sector influences on the  properties of various materials in general
or in  particular on the shear viscosity to entropy density ratio, 
near the critical temperature of the superfluids, might find possible application 
for the future detection experiments of this
mysterious component of the mass of our Universe. Why {\it dark matter} seems to be so important?  
Because  it constitutes almost 24 percent of the mass of the
observable Universe and we practically do not know its provenience. Astronomical observations 
(measurements) enable us to determine the abundance of the {\it dark matter}
with great precision, but we are still lack of the direct experimental confirmation of it, 
different from the astronomical ones mainly based on the gravitational lensing effects. 

{\it Dark matter} influences on the condensed matter systems were intensively 
studied in \cite{nak14}-\cite{pen17}. It turned out that the existence of the {\it dark matter}
sector in AdS/CFT changes some parameters of the systems under considerations. 
These facts are of a great importance from the point of view of the future experiments.

In our considerations we shall use a model of {\it dark matter} sector in which except the ordinary 
Maxwell (Yang-Mills) gauge field we have another gauge field coupled to the latter
and representing the {\it dark matter} sector \cite{bri11}. The applied model is well justified by the astronomical 
observations \cite{ger15}-\cite{massey15b}, as well as, the experimental data connected with 
muon anomalous magnetic moment \cite{muon}. 
Moreover, some experiments looking for {\it dark photon} also confirm  assumed  
 structure of {\it dark matter} sector \cite{afa09}-\cite{red13}. The analysis of the model 
with both both visible and {\it dark matter} may 
constitute new possibilities of the detection of  the {\it dark matter} sector in various 
astronomical and astrophysical processes \cite{reg15}-\cite{nak15aa}.

The motivations of our researches is to find the the viscosity to entropy density ratio universality or its breaking, in the theory of superfluids with {\it dark matter} sector.
The exact form of the universality breaking form constitutes the tantalizing question of our studies. 
It can guide the 'future experiments' which can reveal the structure of {\it dark matter}  if the latter
component of the Universe, e.g., in the form of  clusters~\cite{mun16} can be spotted during 
Earth yearly motion. The 
precise experiments could eventually observe the periodic in time variation of the shear viscosities.
In our paper we analytically prove that the universality of the ratio is not destroyed when 
one considers fluctuations of the metric in the underlying theory. We also achieve
the exact form of the viscosity to entropy density ratio when the $SU(2)$-gauge fields fluctuate. The aforementioned ratio depends on the $\alpha$-coupling constant of ordinary to
{\it dark matter} sector. In this sense it bears the imprint of the {\it dark matter}.

%%%%%%%%%%%%%%%%%%%%%%%%%%%%%%%%%%%%%%%%%%%%%%%%%%%%%%%%%%%%%%%%%%%%%%%%%%%%%%%%%
The paper is organized as follows. In Sec.II we present the holographic set-up. One considers the 
five-dimensional AdS Einstein gravity with two Yang-Mills
fields, ordinary one and the other responsible for the {\it dark matter} sector. In Sec.III the viscosity to entropy density ratio was found for the theory in question, taking into account
only gravitational fluctuations. In Sec.IV we turn on both gravitational and gauge field 
fluctuations responsible for changes of Yang-Mills field. Sec.V is devoted to
summary and discussion of the obtained results.

%%%%%%%%%%%%%%%%%%%%%%%%%%%%%%%%%%%%%%%%%%%%%%%%%%%%%%%%%%%%%%%%%%%%%%%%%%%%%%%%%%%%%%%%%%%%%%%%%%%%%
\section{Holographic setup}

In this section we shall elaborate the five-dimensional AdS Einstein $SU(2)$ gravity with two coupled Yang-Mills fields. The first one is responsible for the ordinary matter,
while the other stands for the {\it dark matter} sector.
The action of the theory is given by
\be
S= \int d^5 x \sqrt{-g}~\frac{1}{\kappa^2} \bigg( R + \frac{12}{L^2} \bigg)  + S_{matter},
\ee
where $S_{matter}$ yields
\be
S_{matter} =  \int d^5 x \sqrt{-g}\bigg[ 
- \frac{1}{4} F_{\mu \nu}{}{}^{(a)} F^{\mu \nu (a)} 
 - \frac{1}{4} B_{\mu \nu}{}{}^{(a)} B^{ \mu \nu (a)}   \\
 -\frac{\alpha}{4} F_{\mu \nu}{}{}^{(a)} B^{\mu \nu (a)} \bigg].
\ee
We have denoted by $\mu,~\nu$ are the five-dimensional spacetime indices, while $(a)$ stands for $SU(2)$ one. 
In what follows we conventionally set the AdS  radius $L=1$. In the above action the parameter $\kappa$ denotes five-dimensional gravity constant, while
$R$  is the Ricci scalar for the considered spacetime.
The Yang-Mills field strengths are given as
\be
F_{\mu \nu}{}{}^{(a)} = \na_\mu A_\nu{}^{(a)} - \na_\nu A_\mu{}^{(a)}  + \ep^{abc} A_\mu{}^{(b)}~ A_\nu{}^{(c)},
\ee
for $A_\beta{}^{(a)}$ and the same expression is valid for $B_\beta{}^{(a)}$.
%%%%%%%%%%%%%%%%%%%%%%%%%%%%%%%%%%%%%%%%%%%%

The model under consideration was widely used in studies of the {\it dark matter} influence on condensed 
matter systems, as a potential guidelines
for the future experiments enabling us to detect this elusive ingredient of the Universe. On the other hand,
the justifications of such kind of models were delivered from the top-down perspective, starting from
the string/M-theory \cite{ach16}. Since the theory in question is a fully consistent quantum theory, 
it guarantees  that any phenomenon described by the top-down theory is physical 
The mixing term of two gauge sectors, called {\it kinetic mixing portals} are typical for states for open 
string theories, i.e., both gauge states are supported by D-branes which are separated in the extra dimensions
(as in supersymmetric Type I, Type IIA, Type IIB models). It results in the existence of massive open strings 
which stretch between two D-branes in question. Massive string/brane states existence 
connects different gauge sectors. They can be also realized by M2-branes wrapped on surfaces which intersect 
two distinct codimension four orbifolds singularities (they correspond (at low energy) to massive 
particles which are charged under both gauge groups).
Of course, there are generalizations of this statement to M, F-theory and heterotic string theory. 

On the other hand, the model with two coupled vector fields, was also implemented in a generalization of p-wave superconductivity, for the holographic model of ferromagnetic 
superconductivity \cite{amo14}.

The equations of motion for the considered theory are provided by the following:
\ben \label{g1}
R_{\mu \nu} &-& \kappa^2~\tT_{\mu \nu}   + 4~g_{\mu \nu} = 0,\\ \label{e1}
\na_\mu B^{\mu \nu (a)} &+& \frac{\alpha}{2}\na_\mu  F^{\mu \nu (a)} + \ep^{abc}~B_\mu{}{}^{(b)}~ B^{\mu \nu (c)}
+ \frac{\alpha}{2} \ep^{abc}~B_\mu{}{}^{(b)}~ F^{\mu \nu (c)} = 0,\\ \label{e2}
\na_\mu F^{\mu \nu (a)} &+& \frac{\alpha}{2}\na_\mu  B^{\mu \nu (a)} + \ep^{abc}~A_\mu{}{}^{(b)}~ F^{\mu \nu (c)}
+ \frac{\alpha}{2}~\ep^{abc}~A_\mu{}{}^{(b)}~ B^{\mu \nu (c)} = 0,
\een
where we have denoted
\be
\tT_{\mu \nu} = T_{\mu \nu} - \frac{1}{3}~T~g_{\mu \nu}.
\ee

In order to simplify the above equations we multiply relation (\ref{e1}) by $\alpha/2$ and extract the term $\frac{\alpha}{2} \na_\mu B^{\mu \nu (a)}$. The second term in the equation (\ref{e2})
is replaced by the aforementioned outcome. The final result may be written as
\ben \label{e3}
\tilde \alpha ~\na_\mu F^{\mu \nu (a)} &-& 
\frac{\alpha}{2}
\ep^{abc}~B_\mu{}{}^{(b)}~ B^{\mu \nu (c)}
- \frac{\alpha^2}{4}\ep^{abc}~B_\mu{}{}^{(b)}~ F^{\mu \nu (c)} \\ \nonumber
&+& \ep^{abc}~A_\mu{}{}^{(b)}~ F^{\mu \nu (c)} +
\frac{\alpha}{2}
\ep^{abc}~A_\mu{}{}^{(b)}~ B^{\mu \nu (c)} = 0,
\een
where $\tilde{\alpha} = 1 - \frac{\alpha^2}{4}$.

The energy momentum tensor in the underlying theory yield
\be
T_{\mu \nu} = T_{\mu \nu}(F) + T_{\mu \nu}(B) + \alpha~T_{\mu \nu}(F,~B),
\ee
and its components connected with the adequate gauge fields are given respectively by
\ben \label{ff1}
T_{\mu \nu}(F) &=& \frac{1}{2}~F_{\mu \ga}{}{}^{(a)}  F_{\nu}{}^{\ga (a)} - \frac{1}{8}~g_{\mu \nu}~F_{\mu \nu}{}{}^{(a)} F^{\mu \nu (a) },\\ \label{bb1}
T_{\mu \nu}(B) &=& \frac{1}{2}~B_{\mu \ga}{}{}^{(a)}  B_{\nu}{}^{\ga (a)} - \frac{1}{8}~g_{\mu \nu}~B_{\mu \nu}{}{}^{(a)} B^{\mu \nu (a) },\\ \label{fb1}
T_{\mu \nu}(F,~B) &=& \frac{1}{2}~F_{\mu \ga}{}{}^{(a)}  B_{\nu}{}^{\ga (a)} - \frac{1}{8}~g_{\mu \nu}~F_{\mu \nu}{}{}^{(a)} B^{\mu \nu (a) }.
\een
Both $SU(2)$ Yang-Mills fields, $A_\mu{}{}^{(b)}$ and $B_\mu{}{}^{(b)}$, are dual to some current operators in the four-dimensional boundary field theory. In 
our considerations we shall assume the following ansatz for $SU(2)$ field components:
\ben  \label{ch}
A &=& \phi(r)~\tau^3~dt + w(r)~\tau^1~dx,\\ \label{ch1}
B &=& \eta(r)~\tau^3~dt.
\een
In the  equations (\ref{ch}) and (\ref{ch1}), the $U(1)$ subgroups of $SU(2)$ one, generated by $\tau^3$ Pauli matrix are 
identified with the electromagnetic $U(1)$-gauge field $\phi(r)$,
while the other $U(1)$ group is bounded with
the {\it dark matter} field $\eta(r)$ which is coupled to the Maxwell one. 
Having non-zero component along $x$-direction, the  gauge boson field $w(r)$ is charged under $A_t^{(3)} = \phi(r)$.
$\phi(r)$ is dual to the chemical potential on the boundary of the AdS spacetime.
As far as $w(r)$ is concerned, it is dual  to $x$-component of a charged vector operator.
In our model the condensation of $w(r)$ field spontaneously break the $U(1)$ symmetry and is subject 
to the superfluid phase transition.  Consequently,  the {\it dark matter} $U(1)$-gauge field has the component 
$B_t^{(3)}$  dual to a current operator on the boundary.

The choice of the gauge field components, described by the relation (\ref{ch}) is the only consistent choice 
of the gauge field components allowing the analytic treatment of the problem in question. The direct calculations reveal that
the $x(1)$ and $t(3)$ components of the equation (\ref{e1}) yield
\ben 
\frac{\alpha}{2}~\na_\mu F^{\mu x (1)} &+& \frac{\alpha}{2}~\ep^{132}~B_t{}{}^{(3)}~F^{tx (2)} = 0,\\ \label{eta}
\na_{\mu} B^{\mu t (3)} &+& \frac{\alpha}{2}~\na_\nu F^{\nu t (3)} = 0,
\een
while the same components of the equation (\ref{e2})  are provided by
\ben
\na_\mu F^{\mu x (1)} &+& \ep^{1 b c}~A_\mu {}{}^{(b)}~F^{\mu x (c)} = 0,\\
\na_\nu F^{\nu t (3)} &+& \frac{\alpha}{2}~\na_\nu B^{\nu t (3)} + \ep^{3 b c}~A_\mu{}{}^{(b)}~F^{\mu t (c)} = 0.
\een
Consequently, the relation  (\ref{e3}) can be written in the form as
\ben \label{x1}
\tilde \alpha ~\na_\mu F^{\mu x (1)} &-& \frac{\alpha^2}{4}~\ep^{132}~B_t^{}{}^{(3)}~F^{tx (2)} + 
\ep^{1 b c} ~A_\mu{}{}^{(b)}~F^{\mu x (c)} = 0,\\ \label{t3}
\tilde \alpha ~\na_\mu F^{\mu t (3)} &+& \ep^{3 b c}~A_\mu{}{}^{(b)}~F^{\mu t (c)} = 0.
\een
In what follows we shall elaborate the $x(1)$ and $t(3)$-components of the aforementioned equation.

The metric ansatz for the AdS charged black brane is assumed to have the general  form
\be
ds^2 = - N(r) \sigma^2(r) dt^2 + \frac{dr^2}{N(r)} + \frac{r^2}{f^4(r)} dx^2 + r^2 f^2(r) ( dy^2 + dz^2 ).
\ee
An asymptotically AdS Reissner-Nordstr\"om black hole, with $w(r) = 0$,  is the known solution of the equations of motion for which one receives
\ben \label{ezero}
\sigma(r) &=& f(r) = 1, \qquad N(r) = r^2 - \frac{m}{r^2} + \frac{ 2 {\tilde \mu}^2 ~\kappa^2~r_+^2}{3 r^4},\\ \label{ezero1}
 \phi(r) &=& {\tilde \mu} \bigg(1 - \frac{r_+^2}{r^2} \bigg).
\een
Because of the fact that $w(r) = 0$, for the above solution, it corresponds to the normal phase of the underlying system.
$\tilde \mu$ denotes a chemical potential of the $SU(2)$ gauge field, ~$r_+$ is the charged black brane 
radius of the event horizon, while $m= r_+^4 + 2~\mu^2~\kappa^2~r_+^2/3$.
The value of the chemical potential in the dual boundary field theory we put equal to four, at the phase transition point, as well as, $r_+=1$ and $\mu = {\tilde \mu}/r_+$,
in order to have connection with the previously obtained results in Einstein and Einstein Gauss-Bonnet AdS gravity.

The relations $x(1)$ and $t(3)$, given by (\ref{x1})-(\ref{t3}), are provided respectively by
\ben
w''(r) &+& w'(r)~\bigg( \frac{1}{r} + \frac{\sigma'(r)}{\sigma(r)} + \frac{N'(r)}{N(r)} + \frac{4 f'(r)}{f(r)} \bigg)
+ \frac{\phi^2(r) w(r)}{\talpha~N^2(r)~\sigma^2(r)}  \\ \nonumber
&-& \frac{\alpha^2}{4~\talpha}~\frac{\eta(r)~\phi(r)~w(r)}{N^2(r)~\sigma^2(r)} = 0,\\
\phi''(r) &+& \phi'(r) \bigg( \frac{3}{r} - \frac{\sigma'(r)}{\sigma(r)} \bigg) - \frac{w^2(r)~f^4(r)~\phi(r)}{\talpha~N(r)~r^2} = 0,
\een
where the prime denotes the derivative with respect to  $r$-coordinate.

On the other hand, the equations of motion for Einstein $SU(2)$-gauge {\it dark matter} system imply
\ben \label{tt}
\frac{3 N'(r)}{r~N(r)} &+& \frac{6 ~\sigma'(r)}{r~\sigma(r)} + \frac{3~N'(r)~\sigma'(r)}{N(r)~\sigma(r)} + \frac{N''(r)}{N(r)} + \frac{2~\sigma''(r)}{\sigma(r)} - \frac{4}{r~N(r)} \\
\nonumber
&-& \kappa^2~
\bigg(
\frac{2~\phi'^2(r)}{3~N(r)~\sigma^2(r)} + \frac{w'^2(r)~f^4(r)}{3~r^2} + \frac{2~w^2(r)~\phi^2(r)~f^4(r)}{3~N^2(r) ~\sigma^2(r)~r^2}
\\ \nonumber 
&+& \frac{2~\eta'^2(r)}{3~N(r)~\sigma^2(r)}  + \frac{2~\alpha~\phi'(r)~\eta'(r)}{3~N(r)~\sigma^2(r)} \bigg)=0, 
\een
\ben \label{xx}
- \frac{1}{r} &+& \frac{3~f'(r)}{r~f(r)} - \frac{f'^2(r)}{f(r)} - \frac{N'(r)}{2~r~N(r)} + \frac{f'(r)~N'(r)}{N(r)~f(r)} - \frac{\sigma'(r)}{2~r~\sigma(r)}
+ \frac{f'(r)~\sigma'(r)}{f(r)~\sigma(r)} \\ \nonumber
&+& \frac{f''(r)}{f(r)} + \frac{2 }{N(r)}  - \kappa^2~\bigg(
\frac{f^4(r)~w'^2(r)}{6~r^2} - \frac{f^4(r)~w^2(r)~\phi^2(r)}{6~r^2~N^2(r)~\sigma^2(r)} +
\frac{\phi'(r)}{12~N(r)~\sigma^2(r)}  \\ \nonumber
&+& \frac{\eta'(r)}{12~N(r)~\sigma^2(r)} + \frac{\alpha~\eta'(r)~\phi'(r)}{12~N(r)~\sigma^2(r)} \bigg) = 0,
\een
\be \label{xxyy}
2 + \frac{r~N'(r)}{N(r)} + \frac{r~\sigma'(r)}{\sigma(r)} - \frac{4~r^2}{N(r)} + 
\kappa^2~
\bigg(
\frac{r^2~\phi'^2(r)}{N(r)~\sigma^2(r)} + \frac{r^2~\eta'(r)}{6~N(r)~\sigma^2(r)} + \frac{\eta'(r)~\phi'(r)~r^2}{N(r) ~\sigma^2(r)} \bigg) = 0.
\ee
It is important to note that the $\phi(r)$ and $\eta(r)$ components of the fields (\ref{ch}) - (\ref{ch1}) are not
independent in the presence of the hair on the considered black brane, i.e., when  $w(r) \ne 0$.

%%%%%%%%%%%%%%%%%%%%%%%%%%%%%%%%%%%%%%%%%%%%%%%%%%%%%%%%%%%%%%%%%%%%%%%%%%
\subsection{Expansion}

As in \cite{bas12}, in order to achieve the corrections of the quantities given in our charge AdS black brane ansatz, one expresses any quantity
in the form provided by
\be
K(r) = K_0(r) + \ep~K_1(r) + \ep^2~K_2(r) + \dots,
\ee
where $\ep = 1 - T/T_c$, ~$T_c$ denotes the critical temperature and $\mid \ep \mid \ll 1$. The above expansion reveals the fact that we shall consider superfluid 
at the temperature close to the critical one.

Next, each of the term $K_a$ is expanded as
\be
K_a(r) = K_{a, 0}(r) + \kappa^2~K_{a,2}(r) + \kappa^4~K_{a, 4}(r) + \dots
\ee
We remark that in the paper \cite{bas12} the expansion has been given in $\kappa/g$, where $g$ is 
the coupling constant of the Yang-Mills field. In our case we put the latter equal to one.
The zeroth-order expansion in terms of $\ep$ is provided by the relations (\ref{ezero}) 
and (\ref{ezero1}), where only $N_0$ contains the correction of $\kappa^2$-order.

Einstein equations in $\ep$-order, imply
\ben
2 N_1(r) &+& r N'_1(r) + \frac{8 \kappa^2}{3 r^4} \bigg(
r^3 \phi'_1(r) - 64 \sigma'_1(r) \bigg) 
+  \frac{4}{3} \alpha \kappa^2 \frac{\eta'_1(r)}{r} = 0,\\ 
\sigma'_1(r)  &=& 0,\\
-2 r N_1(r) &-& N'_1(r) r^2 + 2~f'_1(r)  (5r^4-1) 
- \frac{1}{3 r^2} 64 \alpha^2 ~(r^2+1) f'_1(r) \\ \nonumber
+ 2f''_1(r)(r^4 &-&1)
- \frac{64 \alpha^2}{3 r}f''_1(r) (r^2-1) - \frac{\kappa^2}{3} (8 \phi'_1(r) + 4 \eta'_1(r) ) = 0.
\een
On the other hand, for the gauge fields expanded to the same order one arrives at
\ben
\phi''_1(r) &+& \frac{3}{r} \phi'_1(r) - \frac{8}{r^3} \sigma'_1(r) = 0,\\ \label{w}
w''_1(r) &+& \bigg( \frac{1}{r} + \frac{2(r^4 +1) + r^3 N'_0(r)}{r(r^4 - 1) + r^3~N_0(r)} \bigg) w'_1(r)  \\ \nonumber
&+& \frac{ 16 w_1(r)}{\talpha~\bigg( \frac{N_0(r) r^2}{r^2-1} + r^2 +1 \bigg)^2} = 0.
\een
The forms of the above relations are almost the same as for the Einstein-$SU(2)$-gauge case,
i.e. without dark field. The only difference is in the last term of the equation
(\ref{w}), where the correction of the $\alpha$-coupling constant of the {\it dark matter} is present. As we assume that $\mid \ep \mid \ll 1$, the last term is multiplied by
$1/(1 - \alpha^2/4) \simeq 1+ \alpha^4/4$. It is negligibly small.

Having in mind the leading backreaction correction to the considered line element \cite{bas12}, 
for the needed components of the metric tensor we get
\be
\sigma(r) = 1 - \ep^2~\kappa^2 \frac{2~D^2_1}{9(1+r^2)^3}, \qquad N(r) = r^2 - \frac{1}{r^2} + \frac{32 \kappa^2}{3} \bigg(\frac{1}{r^4} - \frac{1}{r^2} \bigg) + \cO(\ep^2~\kappa^2),
\ee
where $D_1$ is an integration constant. Let us calculate  the Hawking temperature of the considered 
charged AdS black brane. Namely, the standard definition of the black hole temperature 
for the Killing vector field normal to the null hypersurface of the event horizon, 
$\chi_\delta~\chi^\delta = 0$, yields
\be
T = \frac{1}{2 \pi}  \sqrt{- \frac{1}{2} \na_\mu \chi^\nu~\na^\mu \chi_\nu} \mid_{r=1} = \frac{1}{2\pi} \sqrt{ - \frac{1}{4}~g^{rr}~g^{tt}~(g'_{tt})^2} \mid_{r=1}.
\ee
In the case under consideration one obtains
\be
T = \frac{1}{\pi}~\bigg( 1 - \frac{16}{3} \kappa^2 + \frac{17}{1260}~\ep^2~\kappa^2~D^2_1 \bigg) + \cO(\ep^{n>2}~\kappa^{n>2}).
\label{TBH}
\ee
%The critical temperature will be achieved when $\ep=0$, i.e., it has the form
Noting that at $T=T_c$, $\epsilon \equiv 0$ one finds from (\ref{TBH}) the transition temperature 
\be
T_c = \frac{1}{\pi}~\bigg( 1 - \frac{16}{3} \kappa^2 \bigg).
\ee

%%%%%%%%%%%%%%%%%%%%%%%%%%%%%%%%%%%%%%%%%%%%%%%%%%%%%%%%%%%%%%%%%%
\section{$\eta_{yz}/s$ ratio universality}

This section will be devoted to finding the viscosity to entropy density ratio, by means of the Kubo's formula. In order to
perform the task one has to turn on fluctuations of the metric in the underlying theory of gravity, 
as well as, fluctuations of the Yang-Mills fields.
To commence with, we shall first restrict our attention to the gravitational fluctuations along 
the $(y,~z)$-plane. We set the linear corrections to the $(y,~z)$ components of the metric 
tensor $h_{yz}(r)$ in the following form:
\be
h_{yz}(r) = r^2~f^2(r)~\Phi(r,~t).
\ee
Having in mind that to the linear order, the Ricci tensor is given by
\be
R^{(\ep)}_{ab} = \p^c \p_{(b} h_{a)c}  - \frac{1}{2}~\p^c\p_c h_{ab} - \frac{1}{2}~\p_a\p_b h_{m}{}^{m},
\ee
and writing the function $\Phi(r,~t)$ by mean of the Fourier transform
\be
\Phi(r,~t) = \int_{-\infty}^{\infty} d\nu~\Phi_\nu(r)~e^{- i \nu t},
\ee
we conclude that the considered Einstein $SU(2)$-gauge {\it dark matter} equations can be cast in the form as
\ben
\Phi''_\nu(r) &+& \bigg( \frac{N'(r)}{N(r)} + \frac{7}{r} \bigg)\Phi'_\nu(r) + \frac{\nu^2}{N^2(r)~\sigma^2(r)}~\Phi_\nu(r) \\ \nonumber
&+&
\bigg( \frac{8}{r^2} + \frac{2~N'(r)}{r~N(r)}  - \frac{8}{N(r)}
+ \frac{\kappa^2~\phi'^2(r)}{6~ \sigma^2(r)~N(r)} \bigg)\Phi_\nu(r)  = 0.
\een

Because of the fact that the graviton perturbation near the event horizon should be ingoing one, 
\be
\Phi_\nu(r) = \bigg( \frac{N(r)}{r^2} \bigg)^{-\ga}~F(r),
\ee
where $\ga = \frac{i~\nu ~\tT}{4}$ and $\tT$ is provided by
\be 
\tT =\frac{1}{\pi}~\bigg( 1 + \frac{16}{3} \kappa^2 - \frac{17}{1260}~\ep^2~\kappa^2~D^2_1 \bigg) + \cO(\ep^{n>2}~\kappa^{n>2}).
\ee
we arrive at the following relation for $F(r)$:
\ben \nonumber
F''(r) &+& \bigg( (1 - 2\ga)\frac{N'(r)}{N(r)} + \frac{7+2\ga}{r} \bigg)F'(r)  +
\bigg[ \ga^2 \frac{N'^2(r)}{N^2(r)}  \\ \nonumber
&+& \frac{N'(r)}{N(r)}~\bigg( \frac{2 - 5\ga -4 \ga^2}{r} \bigg) 
 - \ga~\frac{N''(r)}{N(r)} + 
 \frac{2 \ga~(2\ga -1) + 8}{r^2} + \frac{14~\ga}{r} \\ \nonumber
 &-& \frac{8}{N(r)} + \frac{\kappa^2~\phi'^2(r)}{6~\sigma^2(r)~N^2(r)} \bigg]~F(r) 
+ \frac{\nu^2}{N^2(r)~\sigma^2(r)}~F(r) = 0.
\een

In the next step we expand  $F(r)$ perturbatively.  Consequently one has
\be
F(r) = F_0(r) + \frac{i \nu}{4} F_1(r) + \cO(\nu^2).
\ee
Then near the boundary of the AdS spacetime, each term $F_a(r)$ will be expanded as follows:
\be
F(\ep,~\kappa^2) = \sum_{a,b =0}^\infty \Phi_{a,~2b}(r)~\ep^a~\kappa^{2b}.
\label{expan}
\ee
Combining the above relation and substitution given by the equation (\ref{expan}), we obtain the following result:
\be
F(r) = \Phi_0(r) +\frac{\Phi_2}{r^4} + \frac{i \nu}{4}~\bigg( \Phi_0(r) +\frac{\Phi_2}{r^4} \bigg) + \cO\bigg(\frac{1}{r^6}\bigg).
\ee
The next task is to calculate the retarded Green function $G_R(\nu,~\vec k =0)$ in the way 
presented in \cite{son02}-\cite{her03}.

\be
G^R_{yz,~yz}(\nu,~\vec k =0) = \frac{1}{2 \kappa^2}~\lim_{r \rightarrow \infty} 
\sqrt{g}~g^{rr}~\frac{F(r)}{\Phi_0~\Phi_2} ~\p_r F(r) = - \frac{2}{\kappa^2} - \frac{i\nu}{\kappa^2}.
\ee

Finally one arrives at the following expression:
\be
\eta_{yz} = - \lim_{\nu \rightarrow 0}~\frac{1}{2~\nu}~
Im(G^R_{yz,~yz}(\nu,~\vec k =0) ) = \frac{1}{2~\kappa^2}.
\ee
Thus in the plane perpendicular to the direction were the symmetry is  broken  the viscosity to entropy density 
ratio has  the standard form
\be
\frac{\eta_{yz}}{s} = \frac{1}{4~\pi}.
\ee
The gravitational modes decouple along the direction $(y,z)$ from the gauge degrees 
of freedom and give rise to the universal result of viscosity to entropy density ratio.
This statement is also valid in the case of {\it dark matter} sector.

One can remark that the gravitational modes decouple from the gauge sectors which are characteristic 
to the low-energy limit of the
heterotic string theory, where one has to do with multiple gauge fields.

%%%%%%%%%%%%%%%%%%%%%%%%%%%%%%%%%%%%%%%%%%%%%%%%%%%%%%%%%%%%%%%%%%%%%%%%
\section{Anisotropy, calculation of $\eta_{xy}/s$ ratio}

In the preceding section we have seen that $h_{yz}$ gravity fluctuations transform as the tensor 
modes of $SO(2)$ symmetry group. They totally
decoupled from any other modes, satisfying a massless scalar field equation.

On the other hand, gravitational modes $h_{xy}$ and fluctuations of $SU(2)$-gauge fields in $y$-direction, 
transform as the vector modes of $SO(2)$ symmetry group.
It suggests that the gravitational fluctuations should couple to the other changes of the fields 
and eventually result in some finite corrections to the shear viscosity
to entropy density ratio. In this section 
we shall concentrate on $(x,~y)$ component of the viscosity to entropy density ratio.
One assumes that the gravitational fluctuations of the form
\be
h_{xy} = r^2~f^2(r)~\Psi(r,~t),
\ee
are present, as well as, Yang-Mills gauge field fluctuations are taken into account. Like in the 
preceding section,
we set all the quantities as Fourier transformed. They imply
\ben
\Psi(r,~t) &=& \int_{-\infty}^{\infty} d\nu~\Psi(r)~e^{- i \nu t},\\
\dA_{y}^{(1)} (r,~t) &=& \int_{-\infty}^{\infty} d\nu~\dA_{y}^{(1)} (r)~e^{- i \nu t},\\
\dA_{y}^{(2)} (r,~t) &=& \int_{-\infty}^{\infty} d\nu~\dA_{y}^{(2)} (r)~e^{- i \nu t},\\
\dB_{y}^{(1)} (r,~t) &=& \int_{-\infty}^{\infty} d\nu~\dB_{y}^{(1)} (r)~e^{- i \nu t},\\
\dB_{y}^{(2)} (r,~t) &=& \int_{-\infty}^{\infty} d\nu~\dB_{y}^{(2)} (r)~e^{- i \nu t}.
\een

The resulting equations of motion are given by
\ben \nonumber
\Psi''(r) &+& \bigg(
\frac{7}{r} + \frac{\sigma'(r)}{\sigma(r)} + \frac{N'(r)}{N(r)} +  \frac{4~f'(r)}{f(r)} \bigg)~\Psi'(r) + \bigg[
\bigg(\frac{2}{r} + \frac{2~f'(r)}{f(r)} \bigg)~\bigg(\frac{3}{r} \\ \nonumber
&+&  \frac{\sigma'(r)}{\sigma(r)} + \frac{N'(r)}{N(r)} \bigg)
+ \frac{2}{r^2} + \frac{8~f'(r)}{r~f(r)} + \frac{2~f'^2(r)}{f^2(r)} + 2 \frac{f''(r)}{f(r)} - \frac{8}{N(r)} \bigg] \\ \nonumber
&+& \frac{\nu^2}{N^2(r)~\sigma^2(r)~r^2~f^2(r)}~\Psi(r) 
- \frac{\kappa^2~\phi'^2(r)}{6~\sigma^2(r)~N(r)}~\Psi(r) \\
&=& \frac{\kappa^2}{r^2~f^2(r)~N(r)}~\bigg( T^{(\ep)}_{xy}(F) 
+ \alpha~T^{(\ep)}_{xy}(F,~B)\bigg), 
\een
where for $T^{(\ep)}_{xy}(F)$ and $T^{(\ep)}_{xy}(F,~B)$ we set the following expressions:
\ben
T^{(\ep)}_{xy}(F)  &=& \frac{1}{2 N(r)~\sigma^2(r)} \bigg(
- w(r) \phi^2(r) \dA^{(1)}_y + i\nu w(r) \phi(r) \dA^{(2)}_y \bigg) \\ \nonumber
&+& \frac{1}{2}~w'(r)~N(r)~\dA^{' (1)}_y,\\
T^{(\ep)}_{xy}(F,~B) &=& \frac{1}{2~N(r)~\sigma^2(r)} \bigg(
- w(r)~\phi(r)~\eta(r)~\dB^{(1)}_y  +  \\ \nonumber
&+& i\nu w(r) \phi(r)~\dB^{(2)}_y \bigg) 
+  \frac{1}{2}~w'(r)~N(r)~\dB^{' (1)}_y.
\een

For the gauge fields in question, they are provided by
\ben \nonumber
\talpha~\dA^{''(1)}_y &+& \talpha~\bigg( \frac{\sigma'(r)}{\sigma(r)} + \frac{1}{r} + \frac{N'(r)}{N(r)} - \frac{2~f'(r)}{f(r)} \bigg)~\dA^{'(1)}_y -
\frac{\talpha~r^2~f^2(r)}{N(r)}~W(r)~\Psi(r) \\ \nonumber
&-& \talpha~w'(r)~f^6(r)~\Psi'(r) + \frac{\phi^2(r) + \talpha~\nu^2}{N^2(r)~\sigma^2(r)}~\dA^{(1)}_y - \frac{i \nu (\talpha +1)~\phi(r)}{N^2(r)~\sigma^2(r)}~\dA^{(2)}_y \\ \nonumber
&-& \frac{\alpha^2}{4} ~\frac{(\eta(r)~\phi(r)~\dA^{(1)}_y - i \nu~\eta(r)~\dA^{(2)}_y)}{N^2(r)~\sigma^2(r)}  \\ \nonumber
&-& \frac{\alpha}{2} ~\frac{(\eta^2(r)~\dB^{(1)}_y - i \nu~\eta(r)~\dB^{(2)}_y)}{N^2(r)~\sigma^2(r)} \\ 
&-& \frac{\alpha}{2} ~\frac{(\phi(r)~\eta(r)~\dB^{(1)}_y - i \nu~\phi(r)~\dB^{(2)}_y)}{N^2(r)~\sigma^2(r)} = 0,
\een

\ben \nonumber
\talpha~\dA^{''(2)}_y &+& \talpha~\bigg( \frac{\sigma'(r)}{\sigma(r)} + \frac{1}{r} + \frac{N'(r)}{N(r)} - \frac{2~f'(r)}{f(r)} \bigg)~\dA^{'(2)}_y 
+ \frac{\phi^2(r) + \talpha~\nu^2}{N^2(r)~\sigma^2(r)}~\dA^{(1)}_y \\ \nonumber
&+& \frac{i \nu (\talpha +1)~\phi(r)}{N^2(r)~\sigma^2(r)}~\dA^{(1)}_y - \frac{i \nu ~w(r)~\phi(r)~f^6(r)}{N^2(r)~\sigma^2(r)} ~\Psi(r) \\ \nonumber
&-& \frac{\alpha^2}{4} ~\frac{(\eta(r)~\phi(r)~\dA^{(2)}_y + i \nu~\eta(r)~\dA^{(1)}_y)}{N^2(r)~\sigma^2(r)} \\ \nonumber
&+& \frac{\alpha}{2} ~\frac{(\eta^2(r)~\dB^{(2)}_y  + i \nu~\eta(r)~\dB^{(1)}_y)}{N^2(r)~\sigma^2(r)} \\ 
&+& \frac{\alpha}{2} ~\frac{(\phi(r)~\eta(r)~\dB^{(2)}_y + i \nu~\phi(r)~\dB^{(1)}_y)}{N^2(r)~\sigma^2(r)} = 0,
\een

%%%%%%%%%%%%%%%%%%%%%%%%%%%%%%%%%%%%%%%%%%%%%%%%%%%%%%%%%%%%%%%%%%%%%%%%%%%%%%%%%%%%%
Because of the bewildering complication of the achieved relations, we assume the situation when only ordinary Yang-Mills gauge field will fluctuate. The {\it dark matter} Yang-Mills 
fields do not fluctuate. Our main idea is
to fluctuate the ordinary Yang-Mills fields, which can be done {\it experimentally} and look for the imprint of the 
{\it dark matter} sector on the viscosity to entropy density ratio.
%%%%%%%%%%%%%%%%%%%%%%%%%%%%%%%%%%%%%%%%%%%%%%%%%%%%%%%%%%%%%%%%%%%%%%%%%%%%%%%%%%%

Having in mind the ingoing boundary conditions of the forms as given in the previous section
\ben
\Psi(r) &=& \bigg( \frac{N(r)}{r^2} \bigg)^{-\ga}~P(r),\\
\dA^{(1)}_y &=& \bigg( \frac{N(r)}{r^2} \bigg)^{-\ga}~J(r),\\
\dA^{(2)}_y &=& \bigg( \frac{N(r)}{r^2} \bigg)^{-\ga}~H(r),
\een

Consequently, taking into account that $\dB_{y}^{(1)} (r,~t) = \dB_{y}^{(2)} (r,~t) =0$  the equations of motion reduces to
\ben \label {epp} \nonumber
P''(r) &+& U(r)~P'(r) + V(r)~P(r) + \frac{\nu^2}{N^2(r)~\sigma^2(r)~r^2~f^2(r)}~P(r) \\
&=& \frac{\kappa^2~\phi'^2(r)}{6~\sigma^2(r)~N(r)}~P(r)
+ \frac{\kappa^2}{r^2~f^2(r)~N(r)}~T^{(\ep)}_{xy}(F),
\een
where we have denoted
\ben
U(r) &=& 2 \bigg( \frac{2~\ga}{r} - \frac{\ga~N'(r)}{N(r)} \bigg) + \frac{7}{r} + \frac{\sigma'(r)}{\sigma(r)} + \frac{N'(r)}{N(r)} + \frac{4~f'(r)}{f(r)},\\ \nonumber
V(r) &=& -\ga~\bigg(\frac{r^2}{N(r)}\bigg) ~\bigg( \frac{7}{r} + \frac{\sigma'(r)}{\sigma(r)} + \frac{N'(r)}{N(r)} + \frac{4~f'(r)}{f(r)}\bigg) + \ga (\ga+1)~\frac{N'^2(r)}{N^2(r)} \\ \nonumber
&-& \frac{4~\ga^2~N'(r)}{r~N(r)} - \frac{\ga~N''(r)}{N(r)} + \frac{2~\ga (2\ga -1)}{r^2} + \frac{2}{r^2} + \frac{8~f'(r)}{r~f(r)} + \frac{2~f'^2(r)}{f^2(r)} \\ 
&+& \frac{2~f''(r)}{f(r)} - \frac{8}{N(r)} +
2\bigg( \frac{1}{r} + \frac{f'(r)}{f(r)} \bigg)~\bigg( \frac{3}{r} + \frac{\sigma'(r)}{\sigma(r)} + \frac{N'(r)}{N(r)} \bigg).
\een
The relevant equation for Yang-Mills - gauge fields are provided by
\ben \label{ejj} \nonumber
J''(r) &+& \bigg[
- 2\ga~\bigg(\frac{r^2}{N(r)}\bigg) ~\bigg( \frac{N'(r)}{r^2} - \frac{2~N(r)}{r^3} \bigg) + \frac{1}{r} + \frac{\sigma'(r)}{\sigma(r)} + \frac{N'(r)}{N(r)} \\ \nonumber
&-& \frac{2~f'(r)}{f(r)} \bigg]~J'(r) 
+ Z(r)~J(r) - w'(r)~f^6(r)~P'(r) \\ \nonumber
&+& \bigg[
- \frac{r^2~f^2(r)~W(r)}{N(r)} \ga~w'(r)~f^6(r)~\bigg(\frac{r^2}{N(r)}\bigg) ~\bigg( \frac{N'(r)}{r^2} - \frac{2~N(r)}{r^3} \bigg) \bigg]~P(r) \\ 
&+& \bigg( - \frac{1}{\talpha}~\frac{i \nu~(\talpha +1)~\phi(r)}{N^2(r)~\sigma^2(r)} + \frac{\alpha^2}{4~\talpha}~\frac{ i\nu~\eta(r)}{N^2(r)~\sigma^2(r)} \bigg)~H(r), 
\een
where we set
\ben \nonumber
Z(r) &=& \ga~(\ga+1)~\bigg(\frac{r^4}{N^2(r)}\bigg) ~\bigg( \frac{N'(r)}{r^2} - \frac{2~N(r)}{r^3} \bigg) - \ga~\bigg(\frac{r^2}{N(r)}\bigg) ~\bigg( \frac{N''(r)}{r^2} \\
&-& \frac{2~N(r)}{r^3} 
+ \frac{6~N(r)}{r^4} - \frac{2~N'(r)}{r^3} \bigg) + \frac{1}{\talpha}~\frac{\phi^2(r) + \talpha~\nu^2}{N^2(r)~\sigma^2(r)} \\ \nonumber
&-& \ga~\bigg(\frac{r^2}{N(r)}\bigg) ~\bigg( \frac{N'(r)}{r^2} - \frac{2~N(r)}{r^3} \bigg)~\bigg( \frac{1}{r} + \frac{\sigma'(r)}{\sigma(r)} + \frac{N'(r)}{N(r)} - \frac{2~f'(r)}{f(r)} \bigg).
\een
For the other component gauge field equations of motion we get
\ben \label{ehh} \nonumber
H''(r) &+& \bigg[ - 2 \ga~\bigg(\frac{r^2}{N(r)}\bigg) ~\bigg( \frac{N'(r)}{r^2} - \frac{2~N(r)}{r^3} \bigg) + 
\frac{1}{r} + \frac{\sigma'(r)}{\sigma(r)} \\ \nonumber
&+& \frac{N'(r)}{N(r)} - \frac{2~f'(r)}{f(r)} \bigg]~H'(r)
+ X(r)~H(r) - \frac{1}{\talpha}~\frac{i \nu ~w(r)~\phi(r)~f^6(r)}{N^2(r)~\sigma^2(r)}~P(r) \\
&+&
\bigg[ \frac{1}{\talpha}~\frac{\talpha~\nu^2 + \phi^2(r)}{N^2(r)~\sigma^2(r)} - \frac{\alpha^2}{4~\talpha}~\frac{i \nu~\eta(r)}{N^2(r)~\sigma^2(r)} \bigg]~J(r),
\een
where we put
\ben \nonumber
X(r) &=& \ga~(\ga+1)~\bigg(\frac{r^4}{N^2(r)}\bigg) ~\bigg( \frac{N'(r)}{r^2} - \frac{2~N(r)}{r^3} \bigg)^2 \\ \nonumber
&-&  
\ga~\bigg(\frac{r^2}{N(r)}\bigg) ~\bigg( \frac{N''(r)}{r^2} - \frac{2~N'(r)}{r^3} + \frac{6~N(r)}{r^4} - \frac{2~N'(r)}{r^3} \bigg) \\ \nonumber
&-& \ga~\bigg(\frac{r^2}{N(r)}\bigg) ~\bigg( \frac{N'(r)}{r^2} - \frac{2~N(r)}{r^3}\bigg) ~\bigg(
\frac{1}{r} + \frac{\sigma'(r)}{\sigma(r)} + \frac{N'(r)}{N(r)} - \frac{2~f'(r)}{f(r)} \bigg) \\
&+& \frac{1}{\talpha}~\frac{i \nu~\phi(r)~(\talpha +1)}{N^2(r)~\sigma^2(r)}  - \frac{\alpha^2}{4~\talpha}~\frac{\eta(r)~\phi(r)}{N^2(r)~\sigma^2(r)} .
\een

In the next step we expand
\ben \label{ep}
P(r) &=& P_0(r) + \frac{i \nu}{4}~P_1(r) + \cO(\nu^2),\\
J(r) &=& J_0(r) + \frac{i \nu}{4}~J_1(r) + \cO(\nu^2),\\ \label{eh}
H(r) &=& H_0(r) + \frac{i \nu}{4}~H_1(r) + \cO(\nu^2),
\een
Just like in the preceding section, using the above relations, one can calculate the retarded Green function bounded with the $(x,~y)$ component of viscosity.
Namely, using the expansion relations given by (\ref{ep})-(\ref{eh}), we substitute them into equations (\ref{epp})-(\ref{ehh}), we find the near-AdS boundary expansions for $SU(2)$ Yang Mills fields.
Because of the fact that $SO(3)$ symmetry is broken spontaneously, the gauge fields should not provide any source to the dual field theory \cite{bas12}. It caused that the components of gauge field which 
fluctuates ought to become normalizable modes of the solutions. As in the later case in order to find the shear viscosity to entropy density ratio we use retarded Green function and take into account
the fall order as $\cO(1/r^4)$. The result is provide by
\be
\frac{\eta_{xy}}{s} = \frac{1}{4~\pi}~\bigg( 1 + \frac{29~\ep^2~\kappa^2}{1792}~(1 - \alpha)~D_1 \bigg).
\ee
On the other hand, taking into account the relations binding the temperature of black object 
in question and the critical one, it can be easily shown that the ratio of viscosity to entropy density
has the temperature dependence as follows:
\be
1 - 4\pi \frac{\eta_{xy}}{s} = (1 - \alpha)~\frac{17690}{15232}~\pi~T_c~\bigg(1 - \frac{T}{T_c} \bigg).
\ee
In \cite{erd12} it was found numerically that the following relation holds
\be
1 - 4\pi\frac{\eta_{xy}}{s} \sim \bigg(1 - \frac{T}{T_c} \bigg)^\beta,
\ee
where $\beta = 1.00 \pm 0.03$. In the case of {\it dark matter} sector, one can observe that the exponent is equal to one and the same form of relation is fulfilled.
Moreover, we have the imprint of the {\it dark matter} by the factor proportional to $(1-\alpha)$. This proportionality factor is smaller than in the case of the ordinary
$SU(2)$ Yang-Mills model of the superfluids. It envisages the imprint of the {\it dark matter} 
sector on the viscosity to entropy density ratio, as well as, on the  critical temperature
one. Moreover,  it happens that {\it dark matter} does not affect the order of the phase transition.
%\red{ what is related to the
% large N limit underlying the AdS/CFT duality.}

%%%%%%%%%%%%%%%%%%%%%%%%%%%%%%%%%%%%%%%%%%%%%%%%%%%%%
\section{Summary and conclusions}
\label{sec:conclusions}

In the paper we have analyzed the problem of the non-universality of the shear viscosity to entropy 
density ratio in AdS Einstein  $SU(2)$-gauge five-dimensional gravity theory.
We considered two Yang-Mills fields, one of them is bounded with the ordinary matter sector, the other 
represents {\it dark matter}. Our main interest will focus on the problem how {\it dark matter}
will modify the KSS bound in the theory under consideration, or can we find some {\it experimental} hints to detect {\it dark matter} by fluctuating the ordinary electrodynamics fields.

We have considered fluctuations of gravitational field and Yang-Mills around the background metric.
The gauge was chosen in such way that both gravitational and $SU(2)$ fluctuations in $r$-direction
were set to be equal to zero. In the anisotropic phase the bulk gravity system has the 
residual symmetry of $SO(2)$ and $Z_2$ types. It turned out that if we consider only $h_{yz}$ fluctuations
of  gravitational field, the modes in question
totally decouple from the other modes and one obtains the universality of the shear viscosity to entropy
density ratio. This situation was known from the previous studies both in Einstein gravity and its generalizations.
We observe no influence of the {\it dark matter} sector on the aforementioned universality.

On the other hand, we elaborate the case when the following fluctuations are present 
$h_{xy}$ and $\dA^{(1)}_y,~\dA^{(2)}_y$, and there are no {\it dark matter} fluctuations. This choice was
caused by the relative simplicity of the underlying equations and possible experiment issues, in which one 
knows how to perform fluctuations in the ordinary electrodynamics. We can imagine 'advanced detector' in a 
huge clump of {\it dark matter}, the 'experiment' will control changes of electrodynamics and the 
ratio of the shear viscosity.

As a result we get the ratio which is modified by the $\alpha$ coupling
constant of the {\it dark matter} to the ordinary one. The influence is  of linear order in $\alpha$.
For positive values of $\alpha$ ($0<\alpha<1$) the ratio is smaller than that obtained 
for the ordinary anisotropic matter, i.e.,
\be
\bigg(\frac{\eta_{xy}}{s}\bigg)_{ordinary~matter} > \bigg(\frac{\eta_{xy}}{s}\bigg)_{dark~matter}.
\ee
We have also established the temperature dependence of the ratio on the {\it dark matter} sector.
Again in the same range of couplings the prefactor is smaller than in the ordinary matter case. Namely
\be
\bigg(1 - \frac{\eta_{xy}}{s}\bigg)_{dark~matter} \sim (1-\alpha)~\bigg(1 - \frac{T}{T_c} \bigg).
\ee
However, the existence of the {\it dark matter} sector does not influence the order of the phase transition.
In both cases ones that the order is equal to one.

We conclude with the remark about the non-universal value of the one of the viscosity tensor components 
to entropy density ratio, in the theory with  spontaneously broken parity~\cite{rog16b}.
The two components in the $(2+1)$-dimensional model are 
called shear and Hall viscosities, respectively.
%%%%%%%%%%%%%%%%%%%%%%%%%%%%%%%%%%%%%
%In Einstein-Maxwell AdS spacetime the condensation of the pseudo scalar field $\theta$ coupled to the gravitational Chern-Simons term
%is essential to the parity breaking effect and consequently non-zero value of the Hall viscosity. Moreover, the consistent behavior at the zero temperature
%is guaranteed by the pseudo scalar potential built of $\theta^2$ and $\theta^4$ terms.  On the contrary, in the case of {\it dark matter}
%sector one needs the spontaneously broken parity by the pseudo scalar hair, the aforementioned potential and the non-zero
%{\it dark matter} gauge field deformation chemical potential. All these factors give us the non-zero value of the Hall viscosity at the boundary of the spacetime in question.
It has been found~\cite{rog16b} that in the theory with Chern-Simons parity breaking term and the condensation of the 
pseudoscalar field, the shear viscosity is unchanged by the {\it dark matter}, while the Hall 
viscosity has been modified, with 
the correction being of a linear order in the coupling constant to {\it dark matter } sector.

%%%%%%%%%%%%%%%%%%%%%%%%%%%%%%%%%%%%%%%%%%%%%%%%%%%%%%%%%%%%%%%%%%%%%%%%%%%%%%%%%%%%%%%%%%%%%%%%
%%%%%%%%%%%%%%%%%%%%%%%%%%%%%%%%%%%%%%%%%%%%%%%%%%%%%%%%%%%%%%%%%%%%%%%%%%%%%%%%%%%%%%%%%%%%%%%%
%\begin{appendix}

%\section{Irred   } 
%\label{irtf}
%\end{appendix}
\begin{acknowledgments}
 MR was partially supported by the grant DEC-2014/15/B/ST2/00089 of the National Science Center
 and KIW by the grant \mbox{DEC-2014/13/B/ST3/04451}.
 \end{acknowledgments}

%%%%%%%%%%%%%%%%%%%%%%%%%%%%%%%%%%%%%%%%%%%%%%%%%%%%%%%%%%%%%%%%%%%%%%%%%%%%%%%%%%%%%%%%%%%%%%%%%%%%%%%
%%%%%%%%%%%%%%%%%%%%%%%%%%%%%%%%%%%%%%%%%%%%%%%%%%%%%%%%%%%%%%%%%%%

%%%%%%%%%%%%%%%%%%%%%%%%%%%%%%%%%%%%%%%%%%%%%%%%%%%%%%%%%%%%%%%%%%%%%%%%%%%%%%%
\end{document}